\newcommand*{\ATLASLATEXPATH}{}
\newcommand{\AtlasCoordFootnote}{
ATLAS uses a right-handed coordinate system with its origin at the nominal interaction point (IP)
in the centre of the detector and the \(z\)-axis along the beam pipe.
The \(x\)-axis points from the IP to the centre of the LHC ring,
and the \(y\)-axis points upwards.
Cylindrical coordinates \((r,\phi)\) are used in the transverse plane,
\(\phi\) being the azimuthal angle around the \(z\)-axis.
The pseudorapidity is defined in terms of the polar angle \(\theta\) as \(\eta = -\ln \tan(\theta/2)\).
Angular distance is measured in units of \(\Delta R \equiv \sqrt{(\Delta\eta)^{2} + (\Delta\phi)^{2}}\).}
 
\documentclass[PAPER, cernpreprint, atlasdraft=true, texlive=2016, UKenglish, LANGEDIT=false]{\ATLASLATEXPATH atlasdoc}
 
\usepackage{\ATLASLATEXPATH atlaspackage}
\usepackage[backref=false]{\ATLASLATEXPATH atlasbiblatex}
 
\usepackage{\ATLASLATEXPATH atlasphysics}
 
\graphicspath{{logos/}{figures/}}
 
\usepackage{ANA-HIGG-2018-58-PAPER-defs}

\AtlasTitle{Search for the Higgs boson decays $H \to ee$ and $H \to e\mu$ in $pp$ collisions at $\sqrt{s} = 13$~\TeV\ with the ATLAS detector}
 
\AtlasAbstract{
Searches for the Higgs boson decays $H \to
ee$ and  $H \to e \mu$ are performed using data
corresponding to an integrated luminosity of \ulumi
collected with the ATLAS detector in $pp$ collisions at
$\sqrt{s} = 13$~\TeV\ at the LHC. No significant signals are observed, in
agreement with the Standard Model expectation. For a
Higgs boson mass of 125~\GeV, the
observed (expected) upper limit at the 95\% confidence level  on the branching fraction $\mathcal{B}(\hee)$
is  \obslimee (\explimee) and on $\mathcal{B}(\hemu)$ is \obslimemu (\explimemu).
These results represent improvements
by factors of about five and six on the previous best limits
on $\mathcal{B}(\hee)$ and $\mathcal{B}(\hemu)$ respectively.
}
 
\author{The ATLAS Collaboration}

\PreprintIdNumber{CERN-EP-2019-184}
 
\AtlasJournalRef{Phys. Lett. B 801 (2020) 135148}
 
\AtlasDOI{10.1016/j.physletb.2019.135148}

\AtlasCoverSupportingNote{Support note}{https://cds.cern.ch/record/2660225/}

\AtlasCoverAnalysisTeam{ Hanna  Borecka-Bielska,  Jan Kretzschmar, Andrew Mehta,  Thomas Neep, Konstantinos Nikolopoulos, Paul Thompson, Russell Turner}

\AtlasCoverEdBoardMember{Bruno~Mansoulie~(chair)}
\AtlasCoverEdBoardMember{Theodota Lagouri}
\AtlasCoverEdBoardMember{Christos Anastopoulos}
 
\AtlasCoverEgroupEditors{atlas-HIGG-2018-58-editors@cern.ch}
 
\AtlasCoverEgroupEdBoard{atlas-HIGG-2018-58-editorial-board@cern.ch}

\hypersetup{pdftitle={ATLAS document},pdfauthor={The ATLAS Collaboration}}
 
\begin{document}
 
\maketitle

\section{Introduction}
\label{sec:intro}

The discovery of a heavy scalar particle by ATLAS and
CMS~\cite{HIGG-2012-27,CMS-HIG-12-028} provided experimental
confirmation of the Englert--Brout--Higgs
mechanism~\cite{Englert:1964et,Higgs:1964ia,Higgs:1964pj,Guralnik:1964eu,Higgs:1966ev,Kibble:1967sv},
which spontaneously breaks electroweak (EW) gauge symmetry and
generates mass terms for the $W$ and $Z$ gauge bosons. In the Standard
Model (SM) the fermion masses are generated via Yukawa interactions.
The Yukawa couplings to third-generation fermions were determined by
measurements of Higgs boson production and decays~\cite{HIGG-2015-07,
CMS-HIG-16-043,HIGG-2017-07,HIGG-2018-04,
CMS-HIG-18-016,CMS-HIG-17-035,HIGG-2018-13}, and found to be in
agreement with the expectations of the SM. However, there
is currently no evidence of Higgs boson decays into first- or
second-generation quarks or leptons.
 
This Letter presents the first ATLAS searches for $H \to ee$ and for the
lepton-flavour-violating decay $H \to e\mu$ using the full Run 2 dataset
of proton--proton ($pp$) collisions at a centre-of-mass
energy of $\sqrt{s}=13$~\TeV, with an integrated luminosity of \ulumi.
The CMS Collaboration has previously
performed searches for $H \to ee$~\cite{CMS-HIG-13-007} and $H \to
e\mu$~\cite{CMS-HIG-14-040} using LHC Run 1 $pp$ data at $\sqrt{s}=8$~\TeV\
corresponding to an integrated luminosity of $19.7~\ifb$.

In the SM the $H \to ee$ branching fraction is given by $G_F m_H
m_e^2/(4\sqrt{2} \pi \Gamma_H) \simeq 5 \times 10^{-9}$, where $m_H$
and $\Gamma_H$ are the Higgs mass and width respectively.
This branching fraction is far below the
sensitivity of the LHC experiments. Contributions from diagrams that
do not depend on the electron Yukawa coupling $Y_{ee}$ and are non-resonant e.g.\ $H\to ee\gamma$, are expected to be
significantly larger, although still much smaller than present sensitivity.
 
The LHC offers the best constraint on
$Y_{ee}$~\cite{Altmannshofer:2015qra}, which may be larger than
predicted by the SM. The SM forbids lepton-flavour-number-violating
Higgs boson decays.  There are strong indirect constraints on the
off-diagonal $Y_{e\mu}$ coupling, the strongest derived from limits on
the branching fraction of $\mu \to e \gamma$ and the electric
dipole moment of the electron~\cite{Harnik:2012pb}.  However, these indirect
constraints assume SM values for the as yet unmeasured
$Y_{ee}$ and $Y_{\mu\mu}$ Yukawa couplings. Searching for $H \to e\mu$
allows $Y_{e\mu}$ to be constrained directly.

Both analyses presented in this Letter closely follow the search for
the SM Higgs boson decay $H \to \mu\mu$~\cite{HIGG-2016-10}. The
signal is separated from the background primarily by identifying a
narrow peak in the distribution of the invariant mass of
the two leptons $m_{\ell\ell}$ corresponding to the mass of the Higgs boson of
$125$~\GeV~\cite{HIGG-2014-14}. The background in the $ee$ search is
dominated by Drell--Yan (DY) $Z/\gamma^*$ production, with smaller
contributions from top-quark pair ($t\bar{t}$) and diboson production
($ZZ$, $WZ$ and $WW$).
In the $e\mu$ search, a much smaller yield of SM background events is
expected. The DY background only contributes through decays of
$Z/\gamma^* \to \tau \tau \to e \nu_\tau \nu_e \mu \nu_\tau \nu_\mu $.
Thus the production of top quarks,
dibosons  (mainly through $WW \to e\nu_e\mu\nu_\mu$),  $W+$jets and
multijet events, with jets misidentified as leptons, are more important
than in the $ee$ search.

\section{ATLAS detector}
\label{sec:detector}
The ATLAS experiment~\cite{PERF-2007-01,ATLAS-TDR-19,PIX-2018-001} at the LHC is a
multipurpose particle detector with a forward--backward symmetric
cylindrical geometry and a near \(4\pi\) coverage in solid
angle.\footnote{\AtlasCoordFootnote} It consists of an inner tracking
detector (ID) surrounded by a thin superconducting solenoid providing a
\SI{2}{\tesla} axial magnetic field, electromagnetic and hadron
calorimeters, and a muon spectrometer.
 
The ID covers the pseudorapidity range
$|\eta| < 2.5$.  It consists of silicon pixel, silicon microstrip, and
transition radiation tracking detectors.  Lead/liquid-argon (LAr)
sampling calorimeters provide electromagnetic (EM) energy measurements
with high granularity.  A steel/scintillator-tile calorimeter
in the central pseudorapidity range $|\eta| < 1.7$ measures the energies of hadrons.
The endcap and
forward regions are instrumented with LAr calorimeters for both the EM and
hadronic energy measurements up to $|\eta| = 4.9$. The muon
spectrometer (MS) surrounds the calorimeters up to $|\eta| = 2.7$ and is based on three
large air-core toroidal superconducting magnets with eight coils each.
The field integral of the toroids ranges between \num{2.0} and
\SI{6.0}{\tesla\metre} across most of the detector. The muon
spectrometer includes a system of precision tracking chambers and fast
detectors for triggering.
 
A two-level trigger system is used to select
events~\cite{TRIG-2016-01}. It consists of a
first-level trigger implemented in hardware and using a subset of the
detector information to reduce the event rate to
\SI{100}{\kHz}. This is followed by a software-based high-level
trigger that employs algorithms similar to those used offline and
reduces the rate of accepted events to \SI{1}{\kHz}.
 
\section{Simulated event samples}
\label{sec:datamc}
 
Samples of simulated signal events with a Higgs boson mass of $m_H =
125$~\GeV\ were generated as described below and processed through the
full ATLAS detector simulation~\cite{SOFT-2010-01} based on
GEANT4~\cite{Agostinelli:2002hh}. Higgs boson production via the
gluon--gluon fusion (\ggF) process was simulated using the POWHEG NNLOPS
program~\cite{Nason:2004rx,Frixione:2007vw,Alioli:2010xd,Campbell:2012am,Hamilton:2012np,Hamilton:2012rf,Hamilton:2013fea,Hamilton:2015nsa}
with the PDF4LHC15 set of parton distribution functions
(PDFs)~\cite{Butterworth:2015oua}.
The Higgs boson rapidity in the simulation was reweighted
to achieve next-to-next-to-leading-order (NNLO) accuracy in QCD~\cite{Catani:2007vq}.
Higgs boson production via vector-boson fusion (VBF) and with an associated
vector boson ($VH$) were generated at next-to-leading-order (NLO) accuracy in QCD using the
POWHEG-BOX
program~\cite{Nason:2009ai,Cullen:2011ac,Luisoni:2013kna}.
The $ZH$ samples were simulated for processes with quark--quark initial states,
and the small contribution from gluon--gluon initial states is accounted for in the normalisation of the $ZH$ cross section.
The parton-level events were processed with
PYTHIA8~\cite{Sjostrand:2014zea} for the decay of the Higgs bosons into
the $ee$ or $e\mu$ final states and to simulate parton showering,
hadronisation and the underlying event,
using the AZNLO set of tuned parameters~\cite{STDM-2012-23}. All
samples were normalised to state-of-the-art predictions using
higher-order QCD and electroweak corrections~\cite{deFlorian:2016spz,
Aglietti:2004nj,Actis:2008ug,Actis:2008ts,Anastasiou:2008tj,Pak:2009dg,Harlander:2009bw,Harlander:2009mq,Harlander:2009my,Actis:2008ug,Anastasiou:2015ema,Anastasiou:2016cez,Dulat:2018rbf,Bonetti:2018ukf,
Ciccolini:2007jr,Ciccolini:2007ec,Bolzoni:2010xr,
Ciccolini:2003jy,Brein:2003wg,Brein:2011vx,Altenkamp:2012sx,Denner:2014cla,Brein:2012ne,Harlander:2014wda,Harlander:2018yio}.
The effects arising from multiple $pp$ collisions in the same or
neighbouring bunch crossings (pile-up) were included in the simulation by
overlaying inelastic $pp$ interactions generated with PYTHIA8 using
the NNPDF2.3LO set of PDFs~\cite{Ball:2012cx} and the A3 set of
tuned parameters~\cite{ATL-PHYS-PUB-2016-017}. Events were reweighted such that the
distribution of the average number of interactions per bunch crossing
matches that observed in data. Simulated events were corrected to
reflect the lepton energy scale and resolution, and trigger,
reconstruction, identification and isolation efficiencies measured in data.

To evaluate the uncertainty in the background modelling in the $ee$
channel, a dedicated fast simulation for the dominant DY background
was used to produce a sample of $10^9$ events, equivalent to 40 times the integrated
luminosity of the data. For this sample, $Z/\gamma^*+(0,1)$-jet events were generated
inclusively at NLO accuracy using POWHEG-BOX~\cite{Barze:2013fru} with the
CT10 PDF set~\cite{Lai:2010vv}. Additional $Z/\gamma^*+2$-jet events
were generated with ALPGEN~\cite{Mangano:2002ea} at leading-order accuracy with
the CTEQ6L1 PDF set~\cite{Pumplin:2002vw}. The events were interfaced
to PHOTOS~\cite{Golonka:2006tw} to simulate QED final-state radiation. The effects of  pile-up
and a fast parameterisation of the response of the detector to electrons and jets, using simple
smearing functions, was then applied to the generated
events.

\section{Event selection}
\label{sec:analysis}
Events are recorded using triggers that require either an isolated
electron or an isolated muon above a transverse momentum ($\pt$) threshold of
$26$~\GeV~\cite{TRIG-2016-01,TRIG-2018-05}. Electrons are reconstructed in the
range $|\eta|<2.47$ from clusters of energy deposits in the
calorimeter matched to a track in the ID~\cite{PERF-2018-01}.
Muons are reconstructed in the range $|\eta|<2.5$ by combining
tracks in the ID either with tracks in the MS or, for $|\eta|<0.1$, with
calorimeter energy deposits consistent with a muon~\cite{PERF-2015-10}.
The electrons and
muons are required to be associated with the primary $pp$ collision
vertex, which is defined as the collision vertex with largest sum of $\pt^2$ of
tracks, and to be isolated from other tracks~\cite{PERF-2018-01, PERF-2015-10}.
Each event must
contain either exactly two electrons or an electron and a muon. One
lepton must have $\pt>27$~\GeV\ to ensure a high
trigger efficiency and the other must be of opposite charge and have
$\pt>15$~\GeV.

Requirements on jets are used in this analysis to suppress background
and define a category that has a high sensitivity to signal produced
in the VBF production mode.  Jets in the range $|\eta|<4.5$ and
$\pt>30$~\GeV\ are reconstructed from energy deposits in the
calorimeter~\cite{PERF-2014-07}, using the anti-$k_t$
algorithm~\cite{Cacciari:2008gp,Fastjet} with a radius parameter of
$0.4$. Tracking information is combined using a multivariate
likelihood to suppressed jets from pile-up interactions~\cite{PERF-2014-03}.
 
Backgrounds with top quarks are suppressed by identifying $b$-hadrons
and neutrinos in the final state. Jets in the range $|\eta|<2.5$
containing $b$-hadrons are identified as $b$-jets using a multivariate
algorithm that uses calorimeter and tracking information~\cite{Aad:2019aic}.
Events are rejected if there
is at least one identified $b$-jet. Different working points are used for
the $ee$ and $e\mu$ channels because the latter has a larger top-quark background.
For the $ee$ ($e\mu$) channel the $b$-jet identification efficiency is about 60\%
(85\%) with a rejection factor of about 1200 (25) for light-flavour
jets~\cite{PERF-2016-05}. Neutrinos produced in semileptonic
top-quark decays escape detection and lead to missing transverse
momentum \etmiss, reconstructed as the magnitude of the vector sum
of the transverse momenta of all calibrated leptons and jets and
additional ID tracks associated with the primary vertex (soft
term)~\cite{PERF-2016-07}. Backgrounds with significant \etmiss{} are
suppressed by requiring $\etmiss/\sqrt{H_\mathrm{T}} < 3.5~(1.75)$~\GeV$^{1/2}$
for the $ee$ ($e\mu$) channel, where $H_\mathrm{T}$ is the
scalar sum of the transverse momenta of leptons and jets and $\sqrt{H_\mathrm{T}}$
is proportional to the \etmiss{} resolution.
 
Background from the process $H\to \gamma \gamma$, where the photons
are misreconstucted as electrons, is studied with simulated
events and found to contribute about $0.07\%$ in the $ee$ channel for
a $H\to ee$ branching fraction at the expected limit. It is therefore
neglected in the rest of the analysis.
 
The search is performed in the range of dilepton invariant mass
$110 < m_{\ell\ell} < 160$~\GeV, which allows the background
to be determined with analytic functions constrained
by the sidebands to either side of the
potential signal.

The event sample passing the basic lepton selection is divided into seven
(eight) categories for the $ee$ ($e\mu$) channel that differ in their
expected signal-to-background ratios, to improve the overall
sensitivity of the search. These categories are based on those used
in Ref.~\cite{HIGG-2016-10}, and are found to provide good sensitivity
in the present analyses.
 
First, a low-\pt lepton category `Low $\pt^\ell$' is defined in the
$e\mu$ channel with events in which the subleading lepton has
$\pt<27$~\GeV. This region has a significant fraction of events in
which either reconstructed lepton is of non-prompt origin or is a
misidentified photon or hadron, hereafter called a fake lepton. These
events are not separated out in the $ee$ channel because the relative
contribution from fake leptons is smaller. A category enriched in
events from VBF production is defined from the
remaining events by selecting those containing two jets with
pseudorapidities of opposite signs, a pseudorapidity separation
$|\Delta \eta_{jj}|>3$ and a dijet invariant mass $m_{jj}>500$~\GeV.
 
Events that fail to meet the criteria of the `Low $\pt^\ell$' and VBF
categories are classified as `Central' if the pseudorapidities of
both leptons are $|\eta^\ell|<1$ or as `Non-central' otherwise. For each of these two
categories, three ranges in the dilepton transverse momentum $\pt^{\ell\ell}$ are considered:
`Low $\pt^{\ell\ell}$' ( $\pt^{\ell\ell} \le 15$~\GeV), `Mid
$\pt^{\ell\ell}$' ($15<\pt^{\ell\ell} \le 50$~\GeV), and `High
$\pt^{\ell\ell}$' ($\pt^{\ell\ell} >50$~\GeV).
These categories exploit differences in the dilepton mass resolution, which is better for more central leptons, as well as differences in the expected signal-to-background ratio between the signal and the background processes as function of dilepton transverse momentum and rapidity.

\section{Signal and background parameterisation}
\label{sec:modelling}
 
Analytic functions are used to describe the $m_{\ell\ell}$
distributions for both the signal and the background. The \hee and \hemu
signals considered are narrow resonances with a mass and
a width set to the SM values of $m_{H}=125$~\GeV\ and $4.1\,\MeV$ respectively.
The observed signal shapes are thus
determined by detector resolution effects and are parameterised
as a sum of a Crystal Ball function ($F_\mathrm{CB}$)~\cite{Oreglia:1980cs} and a Gaussian
function ($F_\mathrm{GS}$) following Ref.~\cite{HIGG-2016-10}:
\begin{equation*}
\begin{split}
P_{\mathrm {S}}(m_{\ell\ell})~=~& f_{\mathrm {CB}} \times {F_\mathrm {CB}}(m_{\ell\ell} | m_{\mathrm {CB}}, \sigma_{\mathrm {CB}}, \alpha, n) \\
& + (1-f_{\mathrm {CB}})\times {F_\mathrm {GS}} \left(m_{\ell\ell} | m_{\mathrm {GS}}, \sigma_{\mathrm {GS}}^{\mathrm S}\right) \,.
\end{split}
\end{equation*}
The parameters $\alpha$ and $n$ define the power-law tail of the $F_\mathrm{CB}$
distribution, while $m_{\mathrm {CB}}$, $m_{\mathrm {GS}}$,
$\sigma_{\mathrm {CB}}$, and $\sigma_{\mathrm {GS}}^{\mathrm S}$
denote the $F_\mathrm{CB}$ mean value, $F_\mathrm{GS}$ mean value, $F_\mathrm{CB}$ width, and $F_\mathrm{GS}$ width
respectively. The relative normalisation between the terms is governed by the parameter $f_{\mathrm {CB}}$. These
parameters are determined by fitting the simulated signal
$m_{\ell\ell}$ distribution in each category.
In the $ee$ ($e\mu)$ channel the signal mass resolution varies between about $2.0\,\GeV$ ($2.3\,\GeV$) for the central and $2.9\,\GeV$ ($3.0\,\GeV$) for the non-central categories.

The background parameterisation for the $ee$ channel follows
Ref.~\cite{HIGG-2016-10} as the background is very similar. The
$m_{ee}$ distributions in each category are described by a sum of a
Breit--Wigner function ($F_\mathrm{BW}$) convolved with a $F_\mathrm{GS}$, and an exponential
function divided by a cubic function:
\begin{equation*}
\begin{split}
P_{\mathrm B}(m_{ee})~=~& f \times [{F_\mathrm {BW}}(m_{ee} | m_{\mathrm {BW}}, \Gamma_{\mathrm {BW}}) \otimes  {F_\mathrm {GS}}(m_{ee} | \sigma_{\mathrm {GS}}^{\mathrm B})] \\
& + (1-f) \times C {\mathrm e}^{A\cdot m_{ee}}/m_{ee}^{3} \,,
\end{split}
\end{equation*}
 
where $f$ represents the fraction of the $F_\mathrm{BW}$ component when each
individual component is normalised to unity and $C$ is a
normalisation coefficient. The $\sigma_{\mathrm {GS}}^{\mathrm B}$
parameter in each category is fixed to the corresponding average
$m_{\ell\ell}$ resolution as determined from simulated signal events.
For all the categories, the $F_\mathrm{BW}$ parameters are fixed to $m_{\mathrm
{BW}}=91.2$~\GeV\ and $\Gamma_{\mathrm {BW}} =
2.49$~\GeV~\cite{Olive:2016xmw}.  The parameters $f$ and $A$ and the overall normalisation are
left free to be determined in the fit and uncorrelated between different
categories.
 
A Bernstein polynomial of degree two is used to parameterise the
$m_{e\mu}$ distribution of the background in each of the eight categories in the $e\mu$ channel, with
parameters uncorrelated across categories. The choice of background function is validated by an F-test considering Bernstein polynomials of first, second and third degree.
 
The signal yield, which is allowed to be positive or negative, is
constrained using separate binned maximum-likelihood fits to the
observed $m_{\ell\ell}$ distributions in the range
$110<m_{\ell\ell}<160$~\GeV\ in the two channels.  The fits are
performed using the sum of the signal and background models (`$S+B$
model') and are performed simultaneously in all the categories.  In
addition to the background-model parameters described earlier, the
background normalisation in each category and the branching fraction
of the signal are free parameters in the fit.

\section{Systematic uncertainties}
\label{sec:systematics}
 
The signal expectation is subject to experimental and
theoretical uncertainties, which are correlated across the categories.
 
The uncertainty in the combined 2015--2018 integrated luminosity is
$1.7\%$~\cite{ATLAS-CONF-2019-021}, obtained using the LUCID\nobreakdash-2
detector~\cite{LUCID2} for the primary luminosity measurements.  Other
sources of experimental uncertainty include the electron and muon
trigger, reconstruction, identification  and isolation
efficiencies~\cite{PERF-2015-10,,PERF-2018-01}, the $b$-jet identification
efficiency~\cite{Aad:2019aic}, the pile-up modelling~\cite{STDM-2015-05}, the determination
of the \etmiss soft term~\cite{PERF-2016-07}, and the jet energy scale
and resolution~\cite{PERF-2016-04}. The uncertainties in the electron energy scale and
resolution~\cite{PERF-2018-01} and in the muon momentum scale and
resolution~\cite{PERF-2015-10} affect the shape of the signal
distribution as well as the signal acceptance.
 
The total experimental uncertainty in the predicted signal yield in
each \ggF category is between 2\% and 3\% for the $ee$ channel
and between 4\% and 6\% for the $e\mu$ channel. It is dominated by the
luminosity, \etmiss soft term and pile-up effects, and the last two contributions
are larger in the $e\mu$ analysis due to the tighter $\etmiss/\sqrt{H_{\mathrm T}}$
requirement. The experimental uncertainty in the VBF category is between 7\%
and 15\% for the $ee$ channel and between 6\% and 22\% for the $e\mu$
channel, due to larger
contributions from the jet energy scale and resolution.

The theoretical uncertainties in the production cross section of the
Higgs boson are taken from Ref.~\cite{deFlorian:2016spz}.  In
addition, theoretical modelling uncertainties affecting the acceptance
for the signals are calculated separately for the \ggF and VBF Higgs
boson production processes in each analysis category.
The uncertainty in the acceptance for the $VH$ process is neglected.
The effects of missing higher-order terms
in the perturbative QCD calculations are estimated by varying the
renormalisation and factorisation scales. For the \ggF process the
uncertainties are approximated as two correlated sources that range from around 1\% to 11\%
for the different analysis categories in both channels.
For the VBF process the uncertainties in the acceptance due to  the
QCD scales are found to be small.
The effects of uncertainties in the parton distribution functions and the value of
$\alpha_\mathrm{S}$ are estimated using the PDF4LHC15
recommendations~\cite{Butterworth:2015oua} and found to be very
small. The uncertainty in the modelling of the parton shower, underlying event, and
hadronisation is assessed by comparing the acceptance of signal
events showered by PYTHIA with that of events showered by
HERWIG~\cite{Bahr:2008pv,Bellm:2015jjp}. The total variations due
to these uncertainties range from less than 1\% to 11\% for the \ggF
signal process and from 1\% to 8\% for the VBF signal process depending on the analysis category.
 
Due to the very different yields and composition of the
backgrounds in the $ee$ and $e\mu$ channels, the potential bias on the
measured signal from the choice of background function is assessed
in different ways. In the $ee$ channel the
$S+B$ fit is repeated using the high-statistics DY-background fast simulation instead
of the data. The number of signal events in each category
obtained from the fit is used as a systematic uncertainty following the
method of Ref.~\cite{HIGG-2012-27}. To be conservative, the maximum
absolute deviation from zero for a signal mass between 120 and 130~\GeV\ is taken.
The uncertainty is treated as uncorrelated between categories. The
background modelling uncertainty is implemented as a set of additional
nuisance parameters acting on the signal normalisation in each
category. The effect of this uncertainty on the expected limit is about 8\%.
In the $e\mu$ channel the background modelling uncertainty is estimated by
changing the fit function to an exponential and evaluating the
difference in signal yield compared with the default  fit to a
sample of simulated background events~\cite{ATL-PHYS-PUB-2017-006, ATL-PHYS-PUB-2017-005, ATL-PHYS-PUB-2018-009}.
The effect of this uncertainty on the expected limit is about $1\%$.

\section{Results}
\label{sec:result}
 
In the $ee$ channel, the observed dielectron mass spectra are divided into 200
$m_{ee}$ bins in each of the seven categories and signal yields are obtained
in a simultaneous maximum-likelihood fit. Confidence intervals are
based on the profile-likelihood-ratio test statistics~\cite{Cowan:2010st},
assuming asymptotic distributions for the test statistics.
The systematic uncertainties affecting the signal normalisation
and shape across categories are parameterised by making the likelihood
function depend on dedicated nuisance parameters, constrained by
additional Gaussian or log-normal probability terms.  The Higgs boson
production cross sections are assumed to be as predicted in the
Standard Model.  The data and expectation for all categories summed
together are shown in Figure~\ref{fig:combined}.  No evidence of the
decay $H\to ee$ is observed.  The best-fit value of the branching
fraction is \obsBFee.  The uncertainty is dominated by the statistical
uncertainty in the data, while the largest systematic contribution
is from the background modelling uncertainty. The observed
(expected) upper limit on the branching fraction, computed using a
modified frequentist CL$_\text{s}$
method~\cite{Cowan:2010st,Read:2002hq}, at the 95\% confidence level,
is found to be \obslimee (\explimee). This result is
a significant improvement on the previous limit by CMS of
$1.9\times 10^{-3}$ based on the Run 1 dataset~\cite{CMS-HIG-13-007}.

In the $e\mu$ channel, a similar fit is performed to the observed
electron--muon mass spectra divided into 50 $m_{e\mu}$ bins in each of
the eight categories. The data and expectation for all
categories summed together are shown in Figure~\ref{fig:combined}. No
evidence of the decay $H\to e\mu$ is observed, with a best-fit value
of the branching fraction of \obsBFemu. The uncertainty is dominated
by the statistical uncertainty in the data, while the largest systematic contribution
is from the Higgs boson production cross-section
uncertainty. The observed (expected) upper limit at the 95\%
confidence level is found to be \obslimemu (\explimemu).  This result is
a significant improvement on the previous
limit by CMS of $3.5\times 10^{-4}$ based on the Run 1
dataset~\cite{CMS-HIG-14-040}.

\begin{figure}[tp]
\centering
\includegraphics[width=0.48\textwidth]{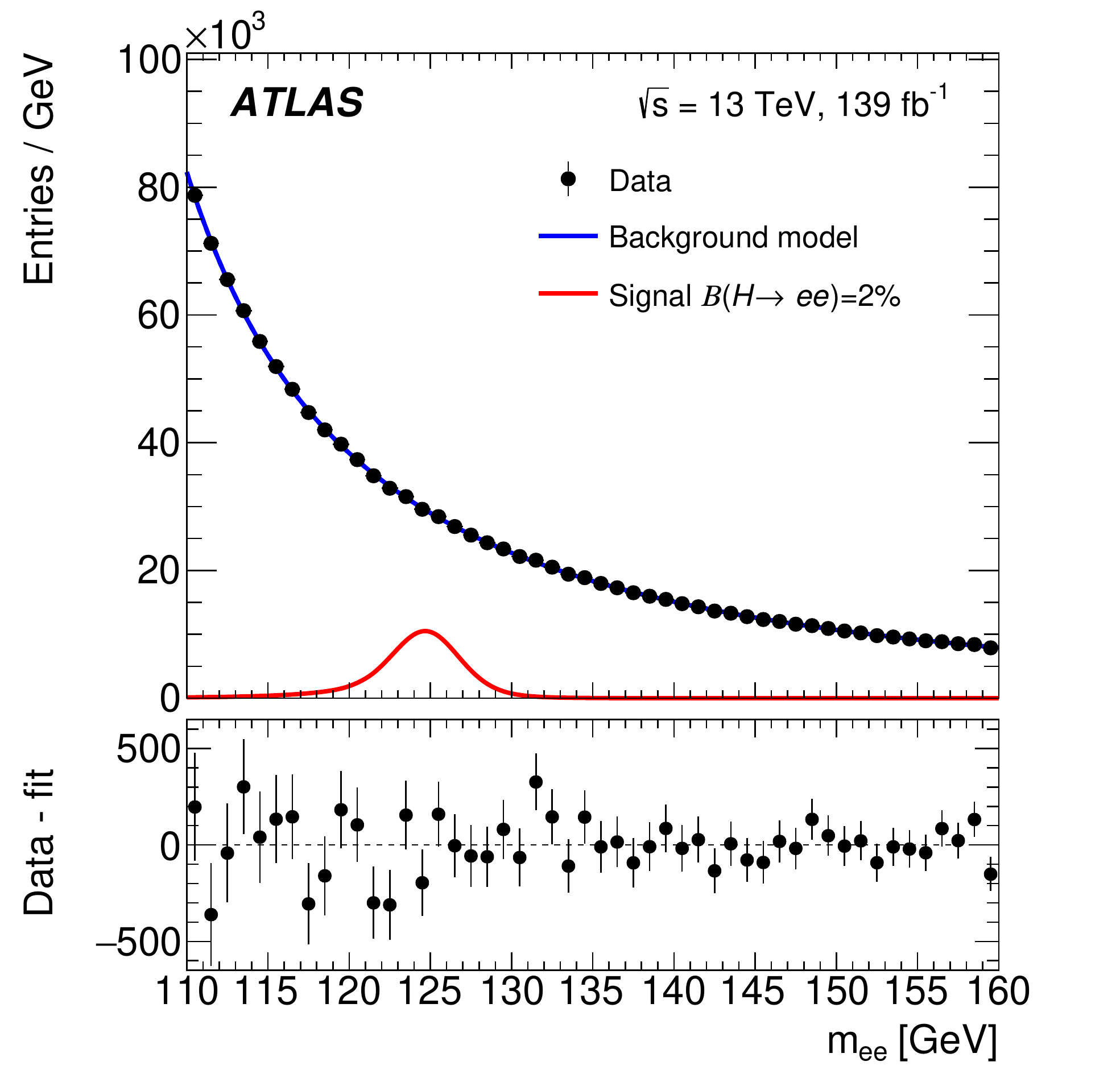}
\includegraphics[width=0.48\textwidth]{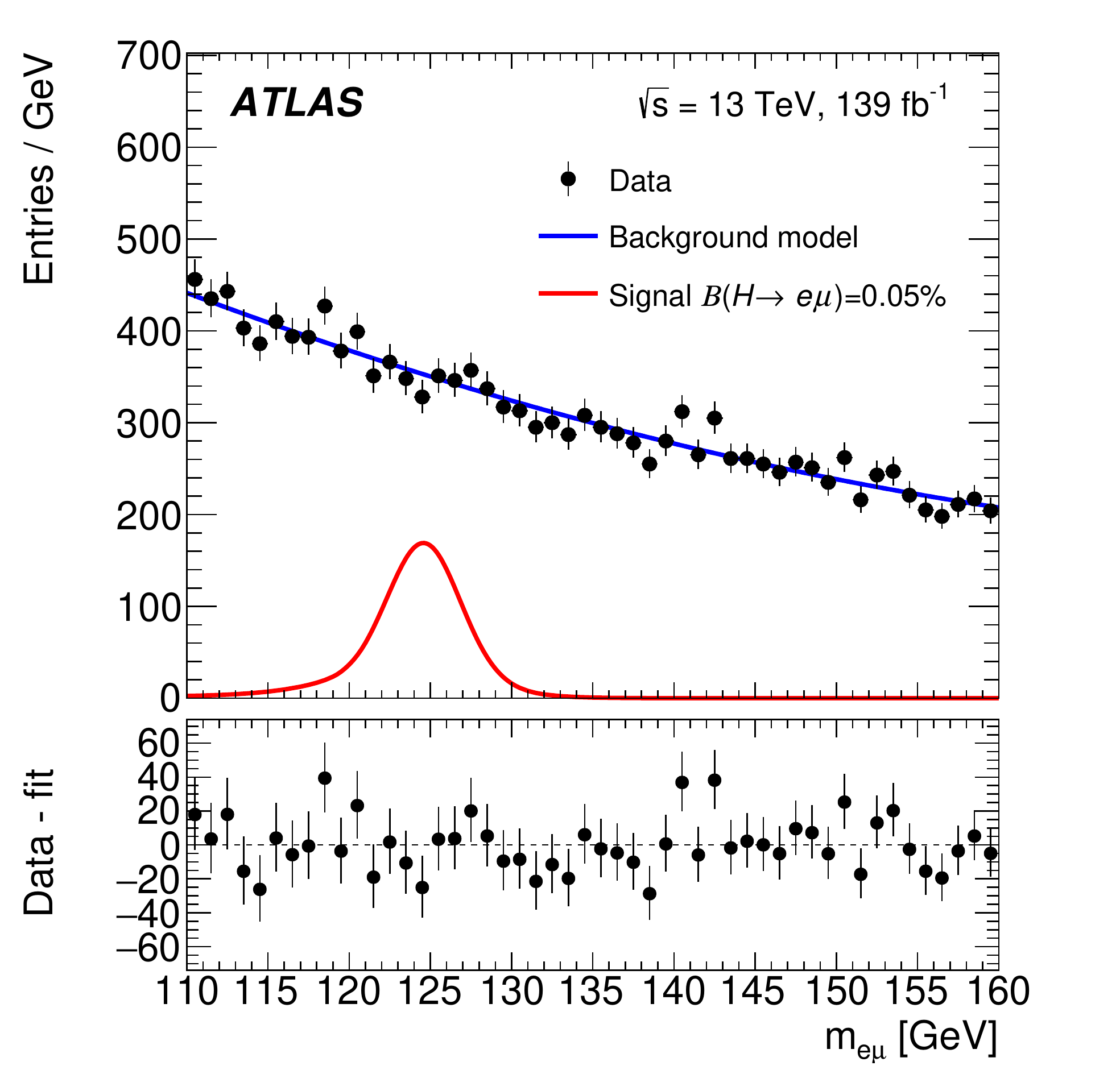}
 
\caption{Dilepton invariant mass $m_{\ell\ell}$ for all categories summed together for the $ee$ channel (left) and
the $e\mu$ channel (right) compared with the background-only model. The
signal parameterisations with branching fractions  set to
$\mathcal{B}(\hee) = 2\%$ and $\mathcal{B}(\hemu) = 0.05\%$ are
also shown (red line). The bottom panels show the difference
between data and the background-only fit.
}
\label{fig:combined}
\end{figure}

\section{Conclusion}
\label{sec:conclusion}
 
Searches are performed for the Higgs boson decays $H \to ee$ and
$H \to e\mu$ using \ulumi of data collected with the ATLAS detector in
$pp$ collisions at $\sqrt{s}=$13~\TeV\ at the LHC. No evidence of
either decay is found and observed (expected) upper limits at the 95\%
confidence level on the branching fractions of \obslimee
(\explimee) for $\mathcal{B}(H \to ee)$ and \obslimemu (\explimemu)
for $ \mathcal{B}(H \to e\mu)$ are obtained for a Higgs boson with mass 125~\GeV.
These are the first such searches made by the ATLAS Collaboration and are considerable
improvements on previous measurements.

\section*{Acknowledgements}
 
 
We thank CERN for the very successful operation of the LHC, as well as the
support staff from our institutions without whom ATLAS could not be
operated efficiently.
 
We acknowledge the support of ANPCyT, Argentina; YerPhI, Armenia; ARC, Australia; BMWFW and FWF, Austria; ANAS, Azerbaijan; SSTC, Belarus; CNPq and FAPESP, Brazil; NSERC, NRC and CFI, Canada; CERN; CONICYT, Chile; CAS, MOST and NSFC, China; COLCIENCIAS, Colombia; MSMT CR, MPO CR and VSC CR, Czech Republic; DNRF and DNSRC, Denmark; IN2P3-CNRS, CEA-DRF/IRFU, France; SRNSFG, Georgia; BMBF, HGF, and MPG, Germany; GSRT, Greece; RGC, Hong Kong SAR, China; ISF and Benoziyo Center, Israel; INFN, Italy; MEXT and JSPS, Japan; CNRST, Morocco; NWO, Netherlands; RCN, Norway; MNiSW and NCN, Poland; FCT, Portugal; MNE/IFA, Romania; MES of Russia and NRC KI, Russian Federation; JINR; MESTD, Serbia; MSSR, Slovakia; ARRS and MIZ\v{S}, Slovenia; DST/NRF, South Africa; MINECO, Spain; SRC and Wallenberg Foundation, Sweden; SERI, SNSF and Cantons of Bern and Geneva, Switzerland; MOST, Taiwan; TAEK, Turkey; STFC, United Kingdom; DOE and NSF, United States of America. In addition, individual groups and members have received support from BCKDF, CANARIE, CRC and Compute Canada, Canada; COST, ERC, ERDF, Horizon 2020, and Marie Sk{\l}odowska-Curie Actions, European Union; Investissements d' Avenir Labex and Idex, ANR, France; DFG and AvH Foundation, Germany; Herakleitos, Thales and Aristeia programmes co-financed by EU-ESF and the Greek NSRF, Greece; BSF-NSF and GIF, Israel; CERCA Programme Generalitat de Catalunya, Spain; The Royal Society and Leverhulme Trust, United Kingdom.
 
The crucial computing support from all WLCG partners is acknowledged gratefully, in particular from CERN, the ATLAS Tier-1 facilities at TRIUMF (Canada), NDGF (Denmark, Norway, Sweden), CC-IN2P3 (France), KIT/GridKA (Germany), INFN-CNAF (Italy), NL-T1 (Netherlands), PIC (Spain), ASGC (Taiwan), RAL (UK) and BNL (USA), the Tier-2 facilities worldwide and large non-WLCG resource providers. Major contributors of computing resources are listed in Ref.~\cite{ATL-GEN-PUB-2016-002}.
 

\printbibliography

\clearpage 
 
\begin{flushleft}
{\Large The ATLAS Collaboration}

\bigskip

G.~Aad$^\textrm{\scriptsize 101}$,    
B.~Abbott$^\textrm{\scriptsize 128}$,    
D.C.~Abbott$^\textrm{\scriptsize 102}$,    
A.~Abed~Abud$^\textrm{\scriptsize 70a,70b}$,    
K.~Abeling$^\textrm{\scriptsize 53}$,    
D.K.~Abhayasinghe$^\textrm{\scriptsize 93}$,    
S.H.~Abidi$^\textrm{\scriptsize 167}$,    
O.S.~AbouZeid$^\textrm{\scriptsize 40}$,    
N.L.~Abraham$^\textrm{\scriptsize 156}$,    
H.~Abramowicz$^\textrm{\scriptsize 161}$
% ...
\bigskip
\\

% ...
$^{40}$Niels Bohr Institute, University of Copenhagen, Copenhagen; Denmark.\\
% ...
$^{53}$II. Physikalisches Institut, Georg-August-Universit\"{a}t G\"ottingen, G\"ottingen; Germany.\\
% ...
$^{70}$$^{(a)}$INFN Sezione di Pavia;$^{(b)}$Dipartimento di Fisica, Universit\`a di Pavia, Pavia; Italy.\\
% ...
$^{93}$Department of Physics, Royal Holloway University of London, Egham; United Kingdom.\\
% ...
$^{101}$CPPM, Aix-Marseille Universit\'e, CNRS/IN2P3, Marseille; France.\\
$^{102}$Department of Physics, University of Massachusetts, Amherst MA; United States of America.\\
% ...
$^{128}$Homer L. Dodge Department of Physics and Astronomy, University of Oklahoma, Norman OK; United States of America.\\
% ...
$^{156}$Department of Physics and Astronomy, University of Sussex, Brighton; United Kingdom.\\
% ...
$^{161}$Raymond and Beverly Sackler School of Physics and Astronomy, Tel Aviv University, Tel Aviv; Israel.\\
% ...
$^{167}$Department of Physics, University of Toronto, Toronto ON; Canada.\\
% ...
$^{a}$ Also at Borough of Manhattan Community College, City University of New York, New York NY; United States of America.\\
$^{b}$ Also at CERN, Geneva; Switzerland.\\
% ...

\end{flushleft}


\end{document}